\documentclass{article}

\usepackage{arxiv}
\usepackage[utf8]{inputenc} 
\usepackage[T1]{fontenc}    
\usepackage{hyperref}       
\usepackage{url}            
\usepackage{booktabs}       
\usepackage{amsfonts}       
\usepackage{microtype}      
\usepackage{amsmath}
\usepackage{bm}
\usepackage{mathtools}
\usepackage{graphicx}
\usepackage{natbib}
\usepackage{multirow}

\DeclarePairedDelimiterX{\norm}[1]{\lVert}{\rVert}{#1}

\title{Gradient-free activation maximization for identifying effective stimuli}

\author{
  Will Xiao \\
  Department of Molecular and Cellular Biology\\
  Harvard University\\
  \texttt{xiaow at fas.harvard.edu} \\
  \And
  Gabriel Kreiman \\
  Center for Brains, Minds, and Machines \\
  Boston Children's Hospital, Harvard Medical School \\
  \texttt{gabriel.kreiman at childrens.harvard.edu} \\
  \And
  {\normalfont Cambridge, MA 02138}
}

\begin{document}
\maketitle

\begin{abstract}
A fundamental question for understanding brain function is what types of stimuli drive neurons to fire. In visual neuroscience, this question has also been posted as characterizing the receptive field of a neuron. The search for effective stimuli has traditionally been based on a combination of insights from previous studies, intuition, and luck. Recently, the same question has emerged in the study of units in convolutional neural networks (ConvNets), and together with this question a family of solutions were developed that are generally referred to as ``feature visualization by activation maximization.''

\vspace{-2.5pt} \setlength{\parindent}{15pt}
We sought to bring in tools and techniques developed for studying ConvNets to the study of biological neural networks. However, one key difference that impedes direct translation of tools is that gradients can be obtained from ConvNets using backpropagation, but such gradients are not available from the brain. To circumvent this problem, we developed a method for gradient-free activation maximization by combining a generative neural network with a genetic algorithm. We termed this method XDream (E\textbf{X}tending \textbf{D}eepDream with \textbf{r}eal-time \textbf{e}volution for \textbf{a}ctivation \textbf{m}aximization), and we have shown that this method can reliably create strong stimuli for neurons in the macaque visual cortex \citep{pxstkm19}. In this paper, we describe extensive experiments  characterizing the XDream method by using ConvNet units as \textit{in silico} models of neurons. We show that XDream is applicable across network layers, architectures, and training sets; examine design choices in the algorithm; and provide practical guides for choosing hyperparameters in the algorithm. XDream is an efficient algorithm for uncovering neuronal tuning preferences in black-box networks using a vast and diverse stimulus space.
\end{abstract}


\section{Introduction}
Consider a typical neuroscience experiment where a visual stimulus is flashed while neural activity is being recorded. Which stimulus properties lead to high firing of the neuron under study? A long-standing idea is that neurons are feature detectors, and their activity indicates the (extent of) presence of their preferred feature in the stimulus. For example, a neuron in primary visual cortex may be tuned to edges oriented at 45 degrees, which means that visual stimuli containing edges of this orientation will trigger stronger responses. The question of preferred stimuli underlies much of our understanding of the visual cortex \citep{hubwie62,pascha02,tsaliv06}.

Recently, the same questions have been asked in the study of artificial neural networks, and much advance has been made in defining the inputs that activate specific units in deep convolutional neural networks (ConvNets). Across the different approaches, one common idea is that of feature visualization by activation maximization: The input image is iteratively optimized to maximize the activation of target units as a way to visualize the features these units represent  \citep{simzis13,olasch17,carola19}.

However, the technique of optimizing stimuli to maximize activation remains relatively little explored in the study of visual neurons (however, see \citealp{yamcha08,carcha11,vazcha14}) due to a key difference between real biological networks and artificial ones: Optimization gradients can be obtained from ConvNets through backpropagation because all the connections and weights are known; in contrast, with current technology, the brain is a black-box where neither connects and weights are known, and thus such gradients cannot be obtained.

We recently introduced an algorithm, XDream (E\textbf{X}tending \textbf{D}eepDream with \textbf{r}eal-time \textbf{e}volution for \textbf{a}ctivation \textbf{m}aximization), that sidesteps the missing gradient problem by combining a non-gradient-based optimization algorithm and a tractable search space \citep{pxstkm19}. Specifically, we used a genetic algorithm to search a space of images parameterized by a generative adversarial network developed by \citet{dosbro16}. We have shown that XDream can create effective stimuli that drive the activity of neurons along the ventral visual stream beyond that elicited by a large set of natural images.

The goal of this paper is to further explain the motivation behind XDream, explore its design choices, and test its performance under a wide range of experimental conditions. It is difficult to thoroughly test XDream directly on biological neurons because recording from neurons is challenging and time-consuming. Instead, in this paper, we use units in ConvNets as \textit{in silico} models of neurons in the ventral visual cortex. Admittedly, ConvNets are only approximate models of the intricacies of ventral visual cortex \citep{ser19}. Nevertheless, ConvNets provide a useful description of primate visual recognition behavior \citep{rajdic18,tankre18} and can explain a significant part of neuronal responses along the ventral stream \citep{yamdic14}. We test multiple ConvNet architectures that are pre-trained for visual recognition tasks. Importantly, we assume no access to any information about the network architecture or weights, thus treating ConvNet units like neurons recorded in an animal. We show that XDream is generalizeable to a wide range of target network architectures and to networks trained on different datasets. Further, we evaluate the performance of XDream when using different initial generations, generative models, and optimization algorithms. We also provide practical guides for choosing hyperparameters in the algorithm. The code for XDream is available at: \url{https://github.com/willwx/XDream/}.

\section{Related Work}
\paragraph{Feature visualization by activation maximization.} Activation maximization is a common approach for understanding the features represented by units in a ConvNet \citep{simzis13,nguclu16,olasch17,olamor18,carola19}. These techniques can only work in networks that provide optimization gradients. In this work, we extend the idea of feature visualization to black-box networks---where we make no assumptions about the architecture or weights---by using gradient-free optimization algorithms.

\paragraph{Finding preferred stimuli of biological neurons.} Evaluating neuronal selectivity in biological systems is traditionally done by a combination of: 1) showing a hand-picked set of stimuli and finding the best ones, and 2) using the former to infer tuning properties of the neuron, then testing images motivated by the hypothesis \citep{hubwie62,brugro81}. Another approach is to use a genetic algorithm to search a parametrically-defined stimulus space \citep{yamcha08,carcha11,vazcha14}. XDream is related to the second approach, but uses a different and arguably more diverse stimulus space. In addition, we frame the approach more broadly and in a modular way, incorporating additional generative models and optimization algorithms as a result. Finally, a recent approach is to fit ConvNet-based models to predict neuron firing, then using standard white-box activation maximization techniques on the ConvNet models \citep{basdic19,waltol18,mal18,abbyu18}. We discuss the relationship between XDream and these approaches below and in \textbf{Discussion}.

\paragraph{Black-box adversarial attack.} The problem of adversarial attack on black box networks is highly related to black-box feature visualization. In black-box adversarial attack, the objective may be to maximize the confidence on a target adversarial class (the ``targeted attack'' scenario; \citealp{ilylin18,chehsi17}). In black-box feature visualization, the objective function is to maximize the activation of a target unit. One difference in our approach is that we use a generative model of images as the domain of optimization. Such generative models are not used in black-box adversarial attacks because the domain is different (minute perturbations). This aside, we borrow concepts and approaches from the body of work on black-box adversarial attack. For example, we evaluated the two algorithms of finite-difference gradient descent and natural evolutionary strategies used in \citet{ilylin18}. In addition, we draw a connection between substitute model-based attack vs. direct attack on one hand, and XDream vs ConvNet-based approaches for activation maximization on the other.

\section{Results}
\subsection{XDream can optimize responses and visualize features without gradients}
\begin{figure}
  \centering
  \includegraphics[width=16.5 cm]{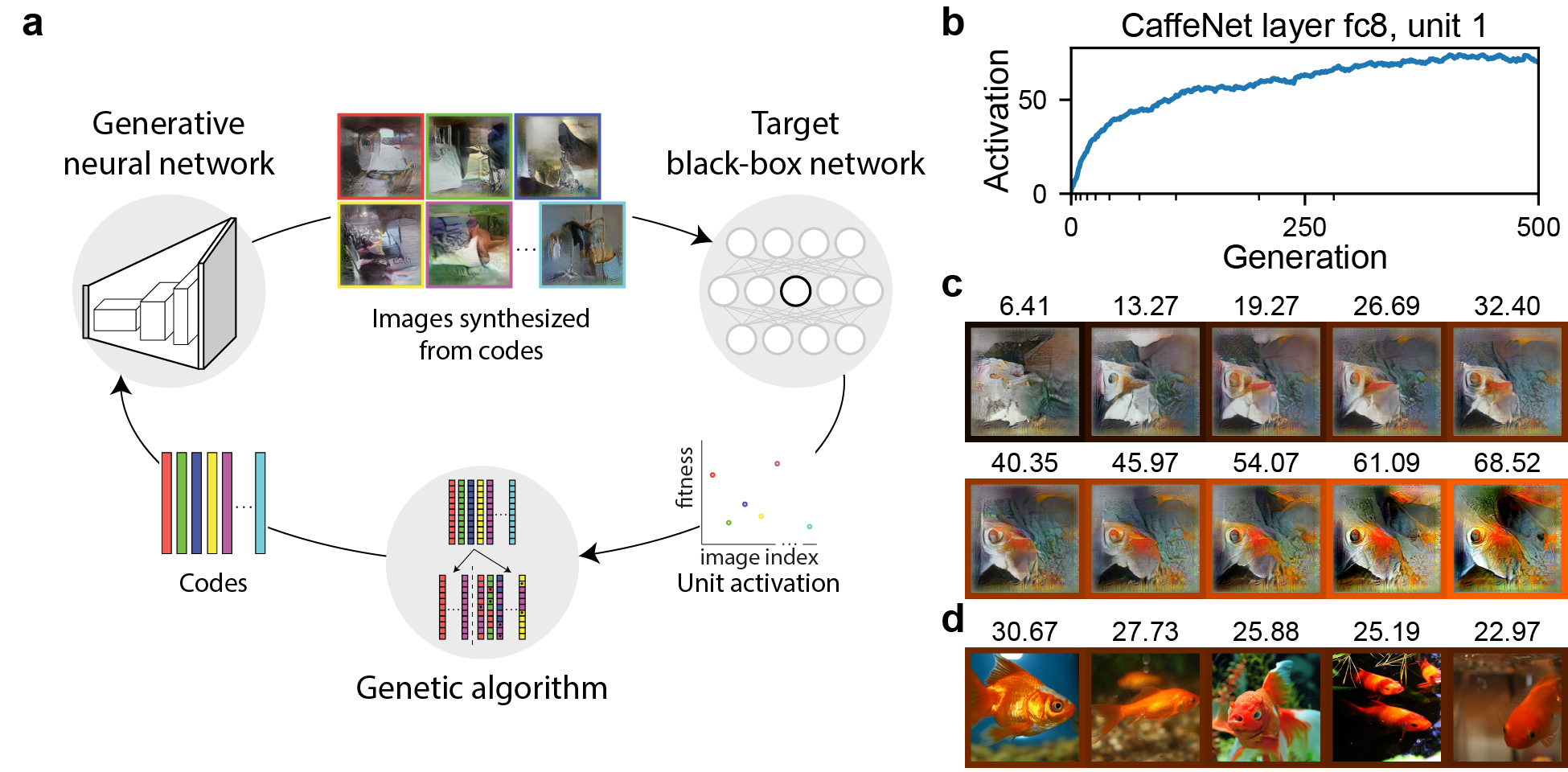}
  \caption{
    \textbf{Overview of the XDream method.}
    \textbf{a)}, XDream combines in an iterative loop a generative model of images (e.g., a generative adversarial network), a target neuron (e.g., a unit in a ConvNet), and a non-gradient-based optimization algorithm (e.g., a genetic algorithm). In each iteration, the optimization algorithm proposes a set of ``codes''---the representation of images in a generative model; the generative model synthesizes the codes into images; the images are evaluated by the target neuron to produce one scalar score per image; finally, the scores are used by the optimization algorithm to propose a new set of codes. Importantly, no gradient information is needed from the neuron.
    \textbf{b,c)}, An example experiment targeting CaffeNet layer fc8, unit 1.
    \textbf{b)}, the mean activation achieved over 500 generations, 20 images each generation (10,000 total image presentations).
    \textbf{c)}, The image obtained at a few example generations indicated by minor x-ticks in \textbf{b)}. The activation to each image is labeled above the image and indicated by the color of the margin.
    \textbf{d)}, The top 5 images among 10,000 random images from ImageNet (ILSVRC12 dataset, $>1.4$ M images). The numeber of images is matched to the number of images presented during optimization. The top image in all of over a million images is shown in \textbf{Figure \ref{fig:nets}b}.
  }
  \label{fig:overview}
\end{figure}

We first show an overview of the XDream approach. XDream has three key components: (i) a generative model of images, (ii) a fitness function given by neural activity, and (iii) an optimization algorithm (\textbf{Figure \ref{fig:overview}a}). An example experiment with unit 1 in layer fc8 of CaffeNet (slightly modified from AlexNet, \citealp{krihin12} by \citealp{don14}) is shown in \textbf{Figure \ref{fig:overview}b}. In a total of 10,000 image presentations, 500 \emph{generations} were presented, each consisting of 20 images. To each presented image, the target unit responds with an \emph{activation} that can be thought of as similar to stimulus-evoked firing rate. The activation of the target unit increased rapidly and saturated at approximately generation 300. \textbf{Figure \ref{fig:overview}c} shows example images at a few generations, log-spaced to show a range of activations. We repeated the experiment with 100 randomly selected units from layer fc8 and obtained strong stimuli in a large majority of cases (CaffeNet output layer in \textbf{Figure \ref{fig:nets}a}).

How strong is the activation elicited by the images generated by XDream? We compared the activation to the best image in the last generation (henceforth referred to as the \emph{optimized image}) and the highest activation to all of the 1,431,167 images in ImageNet (ILSVRC12 dataset; \citealp{rusli15}). For the example unit, the best ImageNet image yielded an activation of 40.55 while the optimized image yielded an activation 72.42 (images shown in \textbf{Figure \ref{fig:nets}}). We refer to the ratio between the activation of the optimized image and activation of the best image in ImageNet as \emph{relative activation}, and we refer to images with relative activation $>1$ (i.e., images that trigger higher activation than any image in ImageNet) as \emph{super stimuli}. In comparison, in a typical neuroscience experiment studying the activity of a neuron in visual cortex, there is a strong limit to how many pictures can be presented while holding stable recordings; typical experiments show somewhere between $n=100$ and $n=10,000$ images \citep{hundic05,tsaliv06,yamdic14}. We estimated the maximum relative activation one can expect to observe by showing this number of natural images, either uniformly randomly sampled from a large image bank or, as is more typical of neuroscience studies, sampled from a few (10) categories (\textbf{Supplementary Section \ref{natmax}}). Since we sample from ImageNet, the relative activation will be $\leq1$ by definition. Even so, for the type of $n$ values explored in neuroscience, the relative activation obtained is well below 1 (\textbf{Figure \ref{fig:natmax}}), suggesting that typical neuroscience experiments do not fully explore the response range of a neuron and are likely to miss stimuli it truly prefers.

\subsection{XDream generalizes across layers, architectures, and training sets}
The generative networks used in XDream were trained on ImageNet using generative features tied to particular layers in CaffeNet (for details about the generative networks, see \citealp{dosbro16}). However, neurons in the brain do not share the same architecture or dataset. Therefore, we tested whether XDream can generalize to units in ConvNets that have different architectures or training sets. We considered 100 units each layer from 4 layers in 6 different network architectures trained on ImageNet: CaffeNet \citep{krihin12,don14}, ResNet-v2 (152 and 269 layers; \citealp{hesun16}),  Inception-v3 \citep{szewoj16}, Inception-v4, Inception-ResNet-v2 \citep{szeale17}; and 1 network with the same architecture as CaffeNet but trained on the Places205 dataset: PlacesCNN \citep{zhooli14}. These network are selected to represent a wide variety of architectures; their internal representations have also been shown to be ``brain-like'' \citep{schdic18}.

\begin{figure}
  \centering
  \includegraphics[width=16.5 cm]{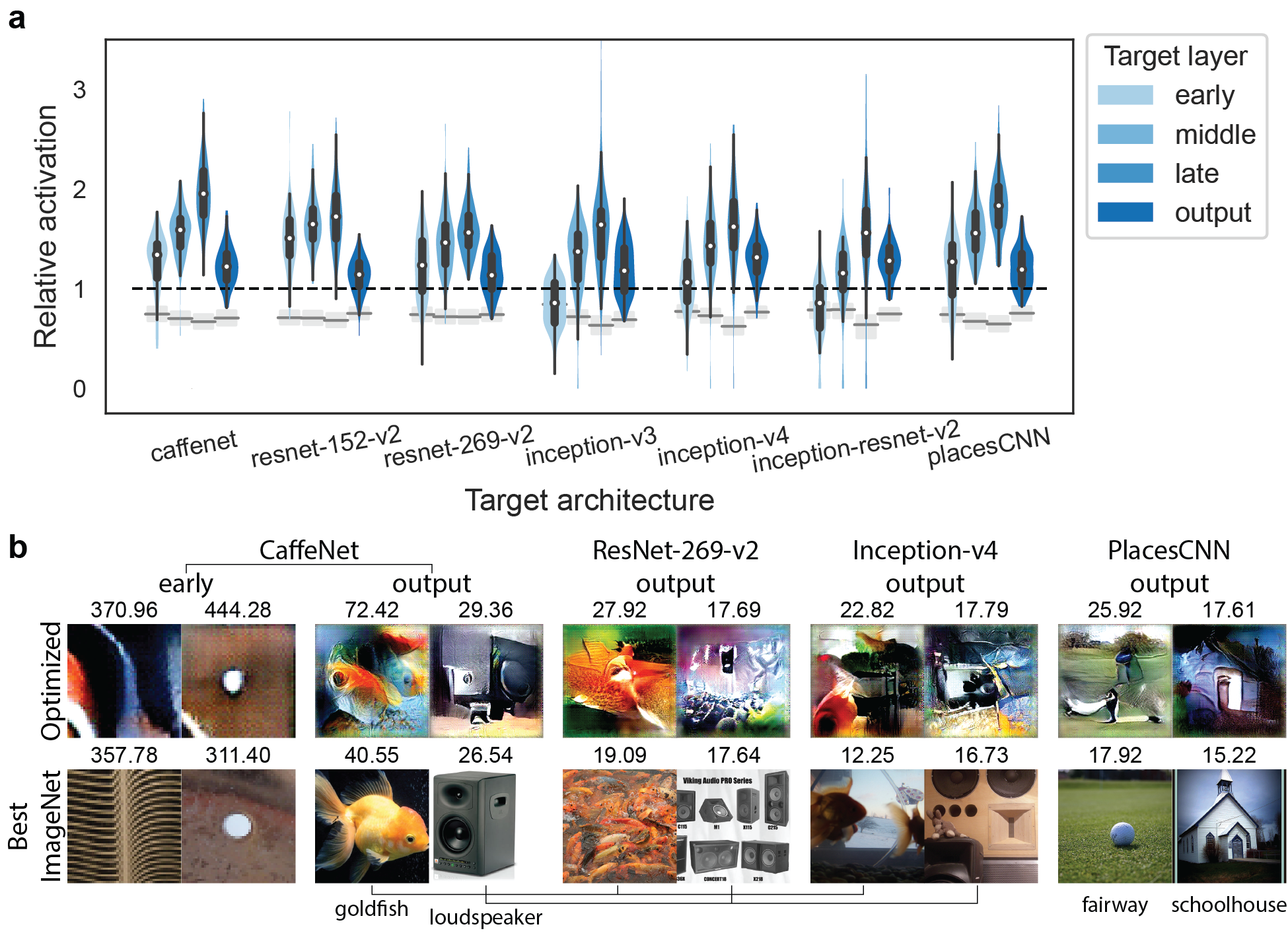}
  \caption{
    \textbf{XDream generalizes across layers, architectures, and training sets.}
    \textbf{a)}, Violin plot showing the distribution of relative activation (activation of optimized stimulus relative to highest activation of $>1.4$ M ImageNet images) over 100 randomly selected units per layer. For each network, we targeted what are roughly early, middle, late, and output layers in the network (the specific layers are indicated in \textbf{Methods}). The contours of the violins show kernel density estimates of the distributions, truncated at min and max observed values. White circles indicate the medians, thick bars indicate the first and third quartiles, and whiskers indicate 1.5$\times$ interquartile ranges. For comparison, grey boxes and lines indicate the distribution (25th-percentile, 75th-percentile, and median) of max relative activation to 10,000 random ImageNet images. The horizontal dashed line indicates 1, corresponding to activation of the best ImageNet image.
    \textbf{b)}, Optimized (top row) and best ImageNet (bottom row) images for a few example units across layers and architectures. Activation values are labeled above the image. For output units, corresponding category labels are shown below. For conv2 units, only the receptive field is shown.
  }
  \label{fig:nets}
\end{figure}

XDream was able to find super stimuli (relative activation $>1$) in the vast majority of units. Compared to a number-matched natural image set, the optimized image exceeded the best natural image in even more units (grey bars in \textbf{Figure \ref{fig:nets}}). Example optimized images for 6 units are shown in \textbf{Figure \ref{fig:nets}b}. For example, for the CaffeNet output unit that detects ``loudspeakers,'' the best ImageNet image is a picture of a loudspeaker, while the optimized image contains features reminiscent of loudspeakers but does not depict a realistic object.

Interestingly, late layers in all networks (and middle layers in most networks) could be driven to higher relative activation than early and output layers, potentially revealing characteristics of the different processing stages. This result is not due to the use of an fc6-based generative network, an alternative hypothesis tested in \textbf{Figure \ref{fig:generators}}.

\subsection{XDream is robust to different initial conditions}
XDream starts with an initial generation. In \textbf{Figure \ref{fig:nets}}, we always initialized with the same set of 20 random image codes, 6 of which are visualized in \textbf{Figure \ref{fig:overview}a}. However, does the choice of initial condition matter?

\begin{figure}
  \centering
  \includegraphics[width=16.5 cm]{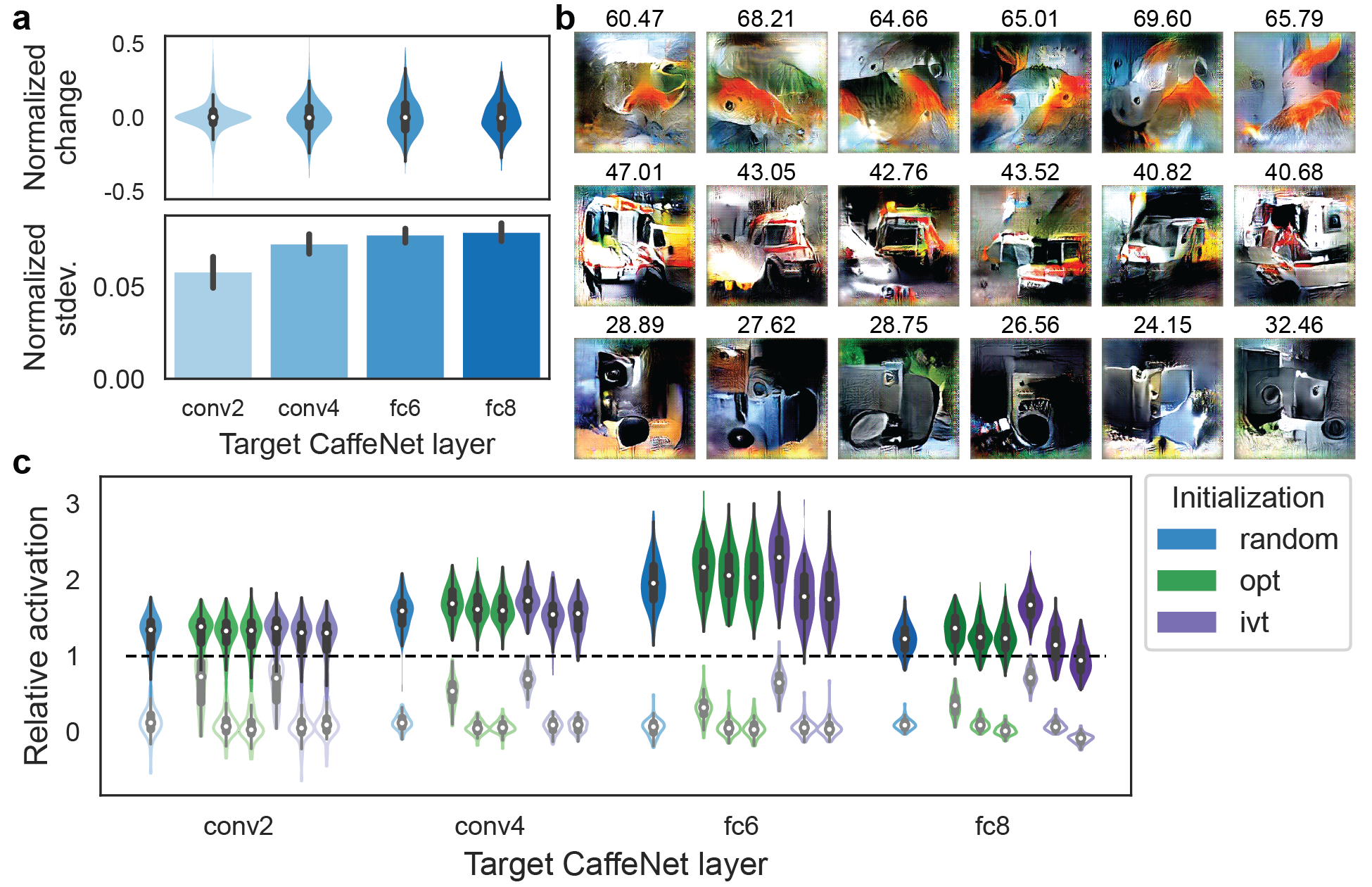}
  \caption{
    \textbf{Comparison of different initializations.}
    \textbf{a,b)}, Effect of using different random initializations.
    \textbf{a)}, Top, distribution of change in optimized activation if one random initialization is changed to one of 10 different random initializations, normalized to optimized activation of the first random initialization. Bottom, standard deviation in optimized activation across 10 different initializations, normalized to the mean. Error bars indicate standard deviation across different units.
    \textbf{b)}, Optimized images from different initializations for 3 example units in the output layer (one unit per row; 6 examples shown). Activation is labeled above each image.
    \textbf{c)}, Good vs. bad initializations. For each target unit, its best, middle, or worst 20 images from ImageNet are used as the initial generation. The images are converted to the image code space using either an optimization method (opt) or an inversion method (ivt; see \textbf{Methods} for details of the encoding methods). Left to right in the ``opt'' and ``ivt'' groups shows initialization with the best, middle, and worst 20 images. Random initialization is shown for comparison. The open and solid violins show the distribution, in the first and last generation respectively, of relative activation over 100 units each layer.
  }
  \label{fig:inits}
\end{figure}

We first asked how much the particular choice of random codes matters. We compared the optimization performance using 10 difference random initializations. In this case, the final activation achieved is highly similar. The standard deviation of optimized activation (i.e., activation of the best image in the final generation) across initializations is lower than 10\% of the activation value, and on average the activation is not expected to change if a different random initialization is used (\textbf{Figure \ref{fig:inits}a}). The resulting optimized images are different on a pixel level, but that is partly expected due to the invariance properties of the units. In comparison, high level features like colors and shapes are preserved (\textbf{Figure \ref{fig:inits}b}).

We next asked whether there are particularly good or bad ways of choosing the initial stimuli. To address this question, we selected, separately for each target unit, the 20 ImageNet images that led to the highest, middle, and lowest activation values, and used those as the initial population (\textbf{Figure \ref{fig:inits}c}). Because the genetic algorithm operates in the image code space, we needed a method to convert an image into an image code. We used two heuristic methods: 1) iteratively optimizing an image code to minimize the pixel-wise difference between the generated and target image (labeled ``opt''); 2) used the CaffeNet fc6 representation of the image as the image code, because the generative network was originally trained to invert this representation (labeled ``ivt''). Please refer to \textbf{Methods} for details of the encoding methods and \textbf{Supplemental Figure \ref{fig:groundtruth}} for example encoded images. Initializing with best vs. worst natural images did not improve the optimized images in the conv2 layer. In progressively higher layers, initializing with better images led to somewhat higher relative activation values both when using the opt and when using the ivt method (\textbf{Figure \ref{fig:inits}c}; \textbf{Table \ref{tab:inits}}). This result indicates that, within the limited number of image presentations, the search is likely to have come closer to the global optimum in earlier layers than it has in later layers. Units in later layers are likely increasingly selective, so it may be more difficult to approach their optimal stimuli. Nevertheless, the improvement relative to the initial generation is not worse when using a worse initialization, suggesting that the optimization is not trapped in local minima when using a worse initialization.

\begin{table}
  \caption{
    \textbf{Effect of using good vs. bad initializations.} For each unit, the 20 best, middle, and worst images from ImageNet, as ranked by that unit, were used as the initial generation. The images were converted to image codes using one of two encoding algorithms, ``opt'' and ``ivt'' (see \textbf{Methods}). The numbers quantify the improvement in median relative activation (across 100 random units each layer) if a better initialization is used (worst $\rightarrow$ middle or middle $\rightarrow$ best). Concretely, it is the linear regression coefficient with the independent variable being \{0,1,2\} for \{worst, middle, best\}, respectively.
  }
  \centering
  \begin{tabular}{rrrrr}
    \toprule
    \multicolumn{1}{c}{ } & \multicolumn{4}{c}{Layer} \\
    \cmidrule(r){2-5}
    Encoding algo & conv2 & conv4 & fc6 & fc8 \\
    \midrule
    opt & 0.010 & 0.037 & 0.047 & 0.056 \\
    ivt & 0.044 & 0.113 & 0.241 & 0.353 \\
    \bottomrule
  \end{tabular}
  \label{tab:inits}
\end{table}

\subsection{Image generators that use high-level representations work equally well}
\begin{figure}
  \centering
  \includegraphics[width=16.5 cm]{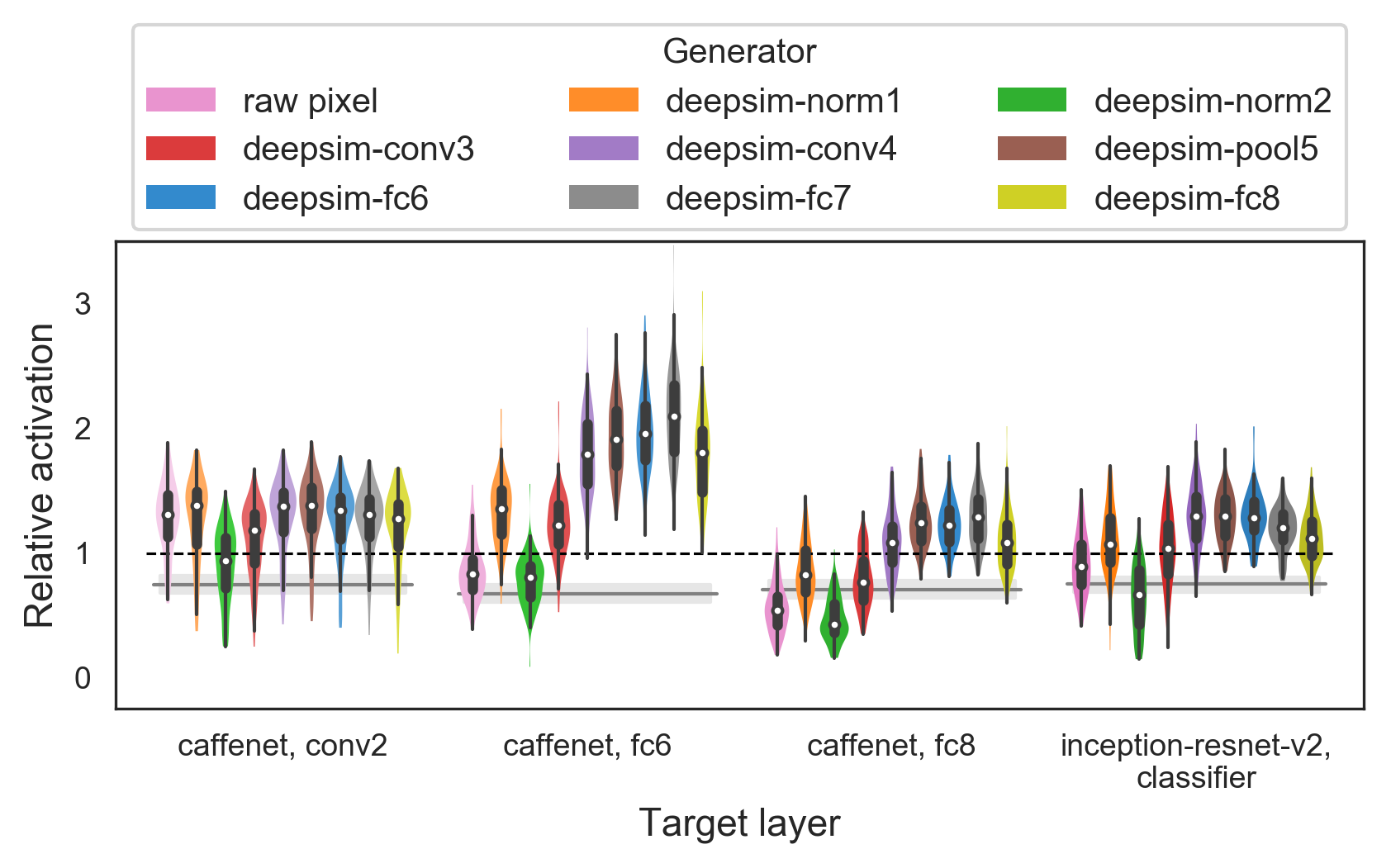}
  \caption{
    \textbf{Comparison of generative models}. We tested each of the family of generative networks from \citet{dosbro16}, as well as a raw pixel-based image representation, as the generative model in XDream. Format of the plot is the same as in \textbf{Figure \ref{fig:nets}a}.
  }
  \label{fig:generators}
\end{figure}

An essential component of XDream is the generative image model. Elsewhere in the paper, we use a generative adversarial network based on CaffeNet fc6 representation \citep{dosbro16}. Here, we consider whether the choice of generative model matters for the performance of XDream and whether the answer depends on the target unit. We hypothesized that generative models based on pixels or low-level representations would not work well, and high-level features may be required for efficient search. To empirically answer these questions, we tested the entire family of DeePSiM generators \citep{dosbro16} trained to invert each layer of CaffeNet (\textbf{Figure \ref{fig:generators}}). The 8 models are of similar depth (11--13 layers) and architecture (feedforward with no bypass or recurrent connections). Notably, models trained on higher layers have more parameters due to having more convolutional filters and, in the case of fc models, having fully-connected layers. In addition to generative neural networks, we also tested a control ``generative model'' where the image is directly parameterized by the flattened pixel array. As targets, we tested early, late, and output layers of CaffeNet, as well as the output layer of Inception-ResNet-v2 to examine cross-model generalization.

Higher layer-based (conv4 and above) generative networks worked similarly well and led to higher activations than early layer-based generative networks (\textbf{Figure \ref{fig:generators}}). The fact that the same generators consistently worked well suggests that the generative model may not need to be tailored to the target unit. One exception was when conv2 units were the targets. In this case, almost all generators worked similarly well. In particular, the pixel-based generator worked poorly for most layers, consistent with expectation, but worked as well as other generators with layer conv2. This result suggests that the selectivity of conv2 units is relatively easy to uncover, consistent with our interpretation for \textbf{Figure \ref{fig:inits}c}. The conv3- and fc8-based generators worked less well than their neighbors. We hypothesize that fc8 representation is task-specific and therefore may not support a good generative model of general images.

Relative activation for CaffeNet fc6 layer was higher than those for conv2 or fc8, as noted already in \textbf{Figure \ref{fig:nets}a}. This observation holds across generative models ranging from conv4-based to fc8-based, consistent with the ideas that the middle layers may intrinsically allow for high relative activation, that it is sufficient for an image generator to have a good dictionary of general features, and that the generator representation need not exactly match that of the target unit.

\subsection{Different optimization algorithms are suited for noiseless and noisy targets}
\begin{figure}
  \centering
  \includegraphics[width=16.5 cm]{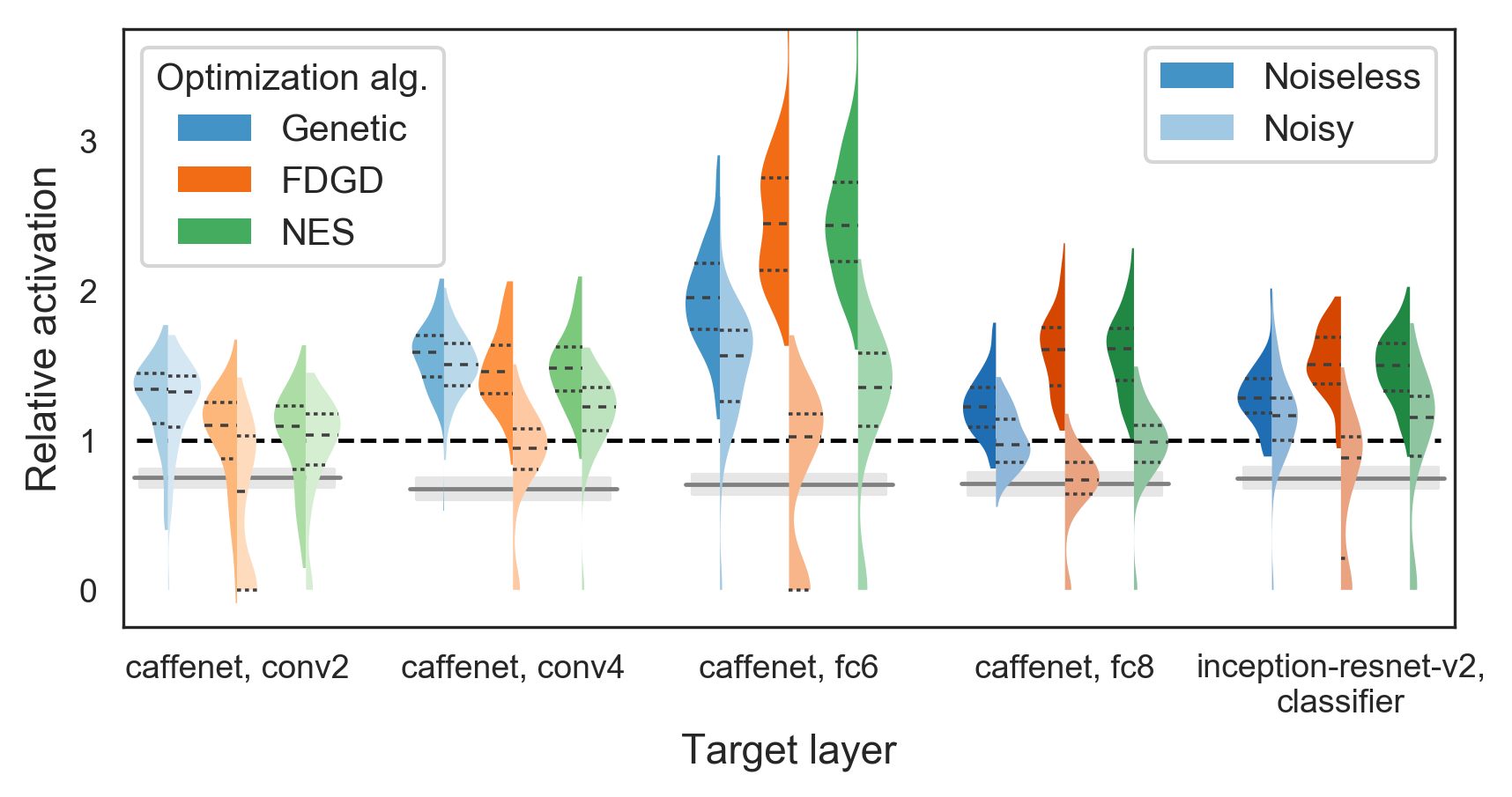}
  \caption{
    \textbf{Comparison of optimization algorithms.} In addition to the genetic algorithm, we evaluated two other gradient-free optimization algorithms: finite-difference gradient descent (FDGD) and Natural Evolution Strategies (NES; \citet{wiesch14}). See \textbf{Methods} for details of the algorithms. Furthermore, we considered optimization performance on noisy units by adding Poisson noise to the activation values. Left and right half of each violin corresponds to noiseless and noisy units, respectively. Dashed lines inside the violin indicate quartiles of the distribution. As in \textbf{Figure \ref{fig:nets}}, grey boxes and grey solid lines indicate the distribution (25th-percentile, 75th-percentile, and median) of expected max relative activation to 10,000 random ImageNet images. The long dashed line indicates 1 and corresponds to the activation to the best ImageNet image.
  }
  \label{fig:optimizers}
\end{figure}

Another essential component of XDream is the optimization algorithm. The optimization algorithm uses the image codes and their corresponding activation values to modify the image codes in order to increase the expected activation. We started by using a genetic algorithm, but other algorithms have been used in related problems \citep{ilylin18}. Therefore, we compared the genetic algorithm to a naïve finite-difference gradient descent algorithm (FDGD; see \textbf{Methods}) and to Natural Evolution Strategies (NES; \citealp{wiesch14}). For CaffeNet conv2 and conv4 layers, all three optimization algorithms yielded comparable results (\textbf{Figure \ref{fig:optimizers}}). For the higher layers both in CaffeNet and Inception-ResNet-v2, the genetic algorithm performed slightly worse.

However, one important property of biological neurons not considered so far is the stochastic nature of their response. That is to say, the same image on repeated presentations can evoke different responses (even though the average response over repetitions can be highly consistent). Noise in objective function evaluations may affect the performance of different optimization algorithms to different extents. Therefore, we compared the optimization algorithms again, this time with a simple model of stochastic neurons: The true activation value is used as the rate of a Poisson process, and the observed values are independent random variables drawn from that Poisson process. Hyperparameters of the optimization algorithms are empirically optimized separately for the noisy case (see \textbf{Section \ref{hyperparams}}). In particular, the algorithms are allowed to present the same image multiple times and average the noisy responses, with the trade-off that fewer unique images can be presented given the same total number of 10,000 presentations. We found repetition (3 repetitions) beneficial only for the FDGD algorithm.

As expected, we found that performance of all algorithms degraded to some extent with noisy target units. Interestingly, the genetic algorithm was at least as good as, and frequently superior to, both alternatives algorithms in all target layers. The NES algorithm performed similarly well as the genetic algorithm in the 3 higher layers (CaffeNet fc6, fc8, and Inception-ResNet-v2 classifier layers). The FDGD algorithm was particularly sensitive to noise, performing the worst in all layers and frequently failed to find good stimuli. It is worth noting that for a Poisson process, the signal-to-noise ratio---in terms of mean over standard deviation---increases as the rate of the Poisson process increases. Because units in lower CaffeNet layers tend to have higher activation values (CaffeNet has no batch normalization), the effective signal-to-noise ratio in our simple model of stochastic neurons may differ from layer to layer. A more realistic future model should take into account more realistic spike numbers and noise distributions.

\subsection{Hyperparameters of the genetic algorithm}
\label{hyperparams}
The optimization algorithms used here all have a number of hyperparameters. Thus, a practical question is what hyperparameter values to use. We consider the genetic algorithm for example and investigate its hyperparameters, briefly described below. For more details on the hyperparameters, please refer to \textbf{Methods}.
\begin{itemize}
    \item Population size ($>$0): the number of individuals in each generation;
    \item Mutation rate (0--1): fraction of components in the image code randomly selected to be mutated in each generation;
    \item Mutation size ($\geq$0): the scale of a zero-centered Gaussian distribution from which mutations are drawn (since the image code is a real-valued vector, mutations are continuous rather than discrete values);
    \item Selectivity ($>$0): during selection, how much to favor high fitness over low fitness. There is no selection when selectivity is zero;
    \item Heritability (0.5--1): what fraction of components in a new image code comes from one of the two parent image codes (since the two parents are arbitrarily ordered, values between 0--0.5 are equivalent to values between 0.5--1).
\end{itemize}

First, we sought a set of useful default values. Since the space of hyperparameters is large and the evaluation of hyperparameter choices computationally expensive (optimization performance has to be evaluated over a set of units), we used a simple greedy algorithm that maximized performance by varying one hyperparameter at a time (see \textbf{Methods}). In this way, we obtained the default hyperparameter values that we used throughout the paper: $\text{population size} = 20, \text{mutation rate} = 0.5, \text{mutation size} = 0.5, \text{selectivity} = 2, \text{heritability} = 0.5$.

\begin{figure}
  \centering
  \includegraphics[width=16.5 cm]{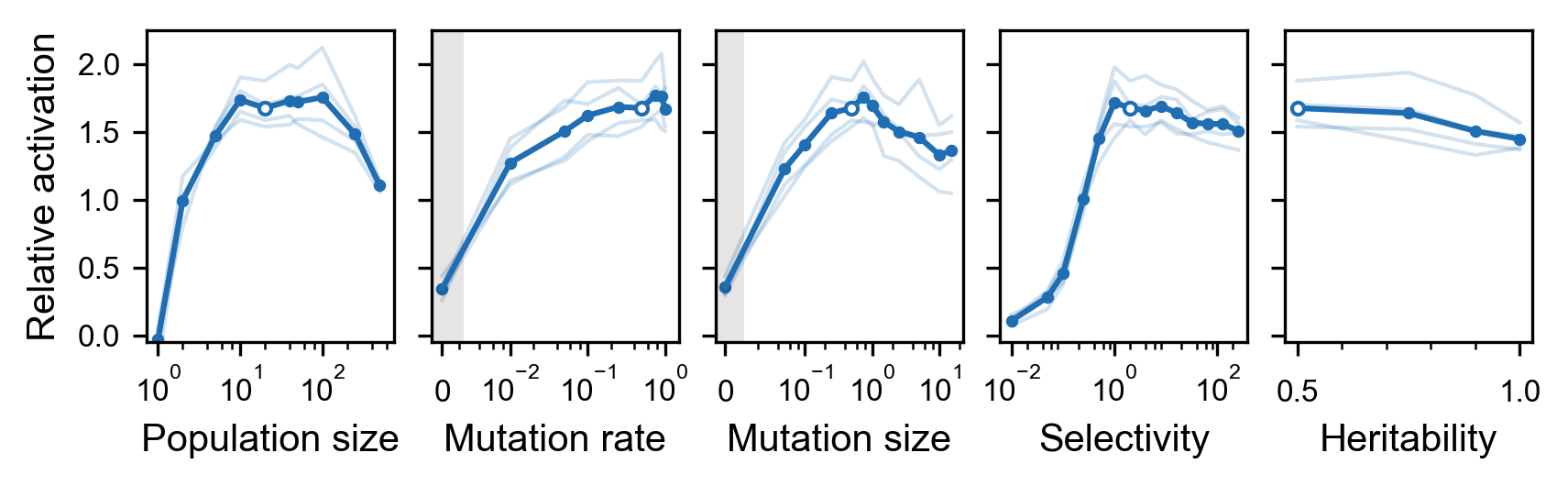}
  \caption{
    \textbf{Comparison of hyperparameters in the genetic algorithm.} In each plot, one hyperparameter is varied while the others are held constant at default values indicated by the open circles. Dots indicate the mean of relative activation across 40 target neurons, 10 neurons each in 4 layers specified in \textbf{Table \ref{tab:itlike}}. Light colored lines indicate the mean across the 10 units within each layer. Light gray shading indicate the linear portion of a symmetrical log plot (in order to show the zero value).
  }
  \label{fig:hyperparams}
\end{figure}

Then, we measured the local fitness landscape around the default values as each parameter is varied independently.  To make the results as transferable to real neurons as possible, we use 40 target units from 4 layers (10 units each) that are found to have similar representation to monkey inferior temporal cortex (\citealp{schdic18}; see \textbf{Methods} for the specific layers). Because these units are different from those used to obtain the default values, the default values may not necessarily be the best values. Nevertheless, the genetic algorithm is robust to a wide range of hyperparameter choices (\textbf{Figure \ref{fig:hyperparams}}). This result is particularly surprising for the selectivity parameter, where values expected to be extremely high are only mildly detrimental. The relative robustness to heritability indicates that recombination is not essential to the performance of the algorithm. Pathologically low parameter values, such as population size of 1, selectivity of 0 (both random diffusion), and mutation rate or mutation size of 0 (both only recombination of extant features), led to poor results as expected.

It is worth remarking that what hyperparameter values are best, and how robust the algorithm is to hyperparameter choices, likely depends on the generative model and the target units. In particular, good hyperparameter values for stochastic target units are expected to differ significantly from the values above, which are tested on noiseless units. Finally, different optimization algorithms having different hyperparameters, so the best hyperparameter values need to be separately explored for each optimization algorithm.

\section{Discussion}
We described the XDream method for discovering preferred images \emph{without assuming any knowledge about the structure or connectivity of the system under study}, and examined its design and performance. An application of this method could be to decipher the inner workings of a vision model for which we lack access to the architecture and weights. The main application we have in mind is elucidating the preferences of neurons in the visual system of a living brain. Neuroscientists mostly lack detailed information about the connectivity and strength of synaptic inputs to neurons in the brain. Consequently, it is difficult to predict which stimulus will produce what activation pattern in the neurons. The choice of stimuli for studying neuron response properties has been guided by a combination of historical data, theory, intuitions, random exploration, and sheer luck. XDream provides a more unbiased and general approach for exploring a vast stimulus space.

How well would the XDream approach extend to the study of biological neurons? Recent work has demonstrated the feasibility of using XDream to uncover the preferences of neurons in different parts of the macaque monkey ventral visual cortex \citep{pxstkm19}. The fact that XDream can generalize across different layers, architectures, and training datasets (\textbf{Figure \ref{fig:nets}}) further suggests that this methodology can be applied to studying different visual areas in the primate brain and potentially the visual system in different animals.

Another approach that has been recently proposed for uncovering visual neuron preference is to optimize white-box models that predict neuron responses \citep{basdic19, waltol18, mal18, abbyu18}. It will be interesting to compare XDream to this alternative approach. A similar comparison has been made in the problem of black-box adversarial attack between the so-called ``substitute network'' approach and what, in comparison, we may call a ``direct'' approach. In substitute network attack, a separate, white-box network is trained to emulate the output of the given black-box network, and the substitute network is attacked in lieu using white-box techniques. In direct attack, no such substitute model is used, and the attack is based on non-gradient-based optimization directly on the black-box target network. It has been argued that direct attack is both free of transferability problems (because no substitute is involved) and more sample efficient \citep{ilylin18,chehsi17}. In this light, to compare the substitute-model approach and the black-box approach for visualizing black-box networks, two key comparisons should be sample efficiency and transferability back to the test case.

The XDream framework appears to be robust and extensible. We examined different ways of initializing the algorithm (\textbf{Figure \ref{fig:inits}}), different generative image models (\textbf{Figure \ref{fig:generators}}), different optimization algorithms (\textbf{Figure \ref{fig:optimizers}}), and different hyperparameters (\textbf{Figure \ref{fig:hyperparams}}). The results suggest that there are ample opportunities for improving each of the components.

For example, in some cases, using good initial conditions could expedite the optimization process (\textbf{Figure \ref{fig:inits}c}). When studying real neurons, investigators may have some prior knowledge about the preferences of a neuron (e.g., from previous experiments or previous runs of XDream). In those cases, initializing with already preferred images may afford some advantage in the final activation value that can be achieved. This result also suggests that XDream does not necessarily reach the \emph{global maximum} in 10,000 image presentations. In other words, there may exist images that trigger higher activations than the optimized stimuli found by XDream. Nevertheless, it is worth emphasizing that number of possible images is combinatorically large, but XDream can identify strong stimuli using just a small number of presentations---stimuli that are much better than can be found by simply presenting more random images (\textbf{Figure \ref{fig:natmax}}). To further investigate how the optimized image may relate to the global maximum, we constructed a simple model where we know the ground truth global maximum by design (\textbf{Supplemental Section \ref{searchability}}). Remarkably, in most cases, XDream was able to essentially uncover the ground truth (\textbf{Figure \ref{fig:groundtruth}}) Thus, XDream may provide a good indication of the globally optimal stimulus.

Another interesting extension is using XDream to optimize different objective functions. The objective function considered in this work is the activation of a single unit; in \citet{pxstkm19} it is the firing rate of single sites recorded by an electrode. In the brain, there are many interesting objective functions, or \emph{neural codes}, that may shed light on different aspects of neural processing. In addition to the firing rate code, it should be possible to use XDream to optimize other neural codes such as local field potential signals, activation of multiple nearby neurons, sparseness of the population representation, synchronous firing of many neurons, etc.

A third interesting future direction is the generative model used. The DeePSiM family of generators \citep{dosbro16} used here were built on ConvNet features as the generative feature space and thereby share a similar internal representation with ConvNets. Because ConvNets have been shown to share similar representations with each other \citep{morben18} as well as with the primate ventral visual stream \citep{yamdic14}, one hypothesis is that DeePSiM generators are uniquely suitable for gradient-free activation maximization because the objective function have simple geometry (e.g., is locally linear) in the generative domain. It is thus an interesting direction for future work whether the XDream approach works equally well with other generative models not based on ConvNet representations, for example BigGAN \citep{brosim18} or compositional pattern producing networks \citep{sta07}.

Finally, although the specific algorithm proposed here focuses on vision, the general framework of XDream should be domain agnostic if given an adequate generative model for the target domain. For example, using a generative model of natural sounds, the same approach here should be transferable to elucidating the stimulus preferences of neurons in the auditory cortex.

In summary, XDream can efficiently explore a vast stimulus space within experimentally accessible time frames, and is less biased than traditional approaches. It has the potential to help elucidate the tuning properties of neurons across brain areas, species, and potentially other sensory modalities and experimental conditions.

\section{Acknowledgements}
This work was supported by the Center for Brains, Minds and Machines funded by NSF STC award CCF-1231216 and also by NIH R01EY026025. We would like to thank Carlos Ponce for comments and discussion.

\section{Methods} \label{methods}
\paragraph{Code availability.}
The code for XDream is available at: \url{https://github.com/willwx/XDream/}.

\paragraph{Generative networks.}
The generative networks are developed by \citet{dosbro16}. The pre-trained models are available on the authors' website: \url{https://lmb.informatik.uni-freiburg.de/people/dosovits/code.html}. The models are used and all experiments done with the caffe library \citep{jiadar14}, but we have converted the models and weights to PyTorch for convenience for future research. Links to the converted models are available on the code repository.

\paragraph{Target architectures and layers.}
We quantified the performance of XDream on multiple layers of several state-of-the-art ConvNet architectures. For each network, we tested what are roughly the early, middle, and late processing stages as well as the output layer. \textbf{Table \ref{tab:targets}} specifies which layers were used in each architecture. One hundred units were randomly selected from each layer except for CaffeNet layer conv2, where all 98 channels are selected. For conv layers, only the center spatial position was selected. All the networks were trained using the ImageNet dataset except PlacesCNN, which was trained using the places dataset \citep{zhooli14}.

\begin{table}
 \caption{\textbf{Target networks and layers.}
 For each network, 4 layers from what is roughly the early, middle, late stages of processing, together with the output layer before softmax, are selected as targets. PlacesCNN has the same architecture as CaffeNet but is trained on the Places-205 dataset \citep{zhooli14}.}
  \centering
  \begin{tabular}{llllll}
    \toprule
    \multicolumn{1}{c}{Network} & \multicolumn{4}{c}{Layers} & As implemented in \\
    \cmidrule(r){2-5}
     & early & middle & late & output & \\
    \midrule
    caffenet & conv2 & conv4 & fc6 & fc8 & \citealp{don14} \\
    \cmidrule(r){6-6}
    resnet-152-v2 & res15\_eletwise & res25\_eletwise & res35\_eletwise & classifier & \multirow{5}{*}{\citealp{alb14}} \\
    resnet-269-v2 & res25\_eletwise & res45\_eletwise & res60\_eletwise & classifier & \\
    inception-v3 & pool2\_3x3\_s2 & reduction\_a\_concat & reduction\_b\_concat & classifier & \\
    inception-v4 & inception\_stem3 & reduction\_a\_concat & reduction\_b\_concat & classifier & \\
    inception-resnet-v2 & stem\_concat & reduction\_a\_concat & reduction\_b\_concat & classifier & \\
    \cmidrule(r){6-6}
    placesCNN & conv2 & conv4 & fc6 & fc8 & \citealp{zhooli14} \\
    \bottomrule
  \end{tabular}
  \label{tab:targets}
\end{table}

\begin{table}
 \caption{\textbf{Inferior temporal cortex-like layers.} From each layer, 10 units are randomly selected and used in hyperparameter evaluation (\textbf{Figure \ref{fig:hyperparams}}).}
  \centering
  \begin{tabular}{ll|ll}
    \toprule
    Network & Layer & Network & Layer \\
    \midrule
    caffenet & pool5 & resnet-101-v2 & res32\_eletwise \\
    placesCNN & pool5 & densenet-169 & concat\_5\_31 \\
    \bottomrule
  \end{tabular}
  \label{tab:itlike}
\end{table}

\paragraph{Optimization algorithms.}
An optimization algorithm in the context of XDream is a function that iteratively: proposes a set of image codes $\bm{c}_i,i=1,\dots,n$ (float vectors), or \emph{codes} for short; then, uses their corresponding \emph{fitness} values $y_i,i=1,\dots,n$ (scalar activations by a target unit to the image associated with each code) to propose a new set of codes expected to have higher fitness.

\vspace{-5pt} \setlength{\parindent}{15pt}
The genetic algorithm works as follows: Each generation consists of $n$ codes, where $n$ is the population size parameter. Their corresponding fitness values $y_i,i=1,\dots,n$ are transformed into probability weights $w_i=\exp(\left(y_i-\min_i(y_i)\right)/k)$, where $k=\text{stdev}_i(y_i)/s$ is analogous to temperature in the Boltzmann equation and $s$ is the selectivity parameter (higher $s\sim$ lower temperature $\sim$ high fitness is more favored). To create each code in the next generation (a progeny), two codes (parents) are drawn with the probability for each code to be drawn equal to $p_i=w_i/\sum_iw_i$ (note that the two parents do not have to be distinct). A random fraction $h$ of vector components in the progeny is drawn from one parent and $(1-h)$ from the other, where $h$ is the heritability parameter. Finally, a fraction $r$ of the components in each progeny is subject to mutation drawn from a zero-centered Gaussian of scale $\sigma$; $r$ is the mutation rate parameter and $\sigma$ the mutation size parameter.

\vspace{-5pt} \setlength{\parindent}{15pt}
The finite-difference gradient descent algorithm works as follows: A set of $2n$ \emph{sample} codes $\bm{c}_{i,\pm},i=1,\dots,n$ is proposed around the current \emph{center} code $\bm{c}_0$ by adding to it zero-centered Gaussian perturbation $\bm{\delta}_i$ of scale $\sigma$, where $\sigma$ is the search radius parameter. The samples are antithetic, meaning that $\bm{c}_{i,\pm}=\left(\bm{c}_0\pm\bm{\delta}_i\right)$. The gradient estimate is then $\Delta\bm{c}_0=\sum_i\Delta y_i\bm{\delta}_i/\norm{\bm{\delta}_i}^2$, where $\Delta y_i=\left(y_{i,+}-y_{i,-}\right)$ and $\norm{\cdot}$ is the L1 norm. The new center is then $\bm{c'}_0=\left(\bm{c}_0+\eta\Delta\bm{c}_0\right)$, where $\eta$ is the learning rate parameter.

\vspace{-5pt} \setlength{\parindent}{15pt}
Natural evolution strategies aim to maximize not $y=f(\bm{c}_0)$ at a center code $\bm{c}_0$, but the expectation $\mathbb{E}_{\pi(\bm{c})}\left[f(\bm{c})\right]$ over a search distribution $\pi(\bm{c})$. We refer the reader to \citet{wiesch14} for motivation and derivation. The implementation is as follows: The search distribution is a Gaussian of scale $\sigma$  around the current center $\bm{c}_0$. A set of $2n$ antithetic samples, $\bm{c}_{i,\pm}, i=1,\dots,n$, is proposed as above. The gradient estimate for the center is  $\Delta\bm{c}_0=\frac{1}{n\sigma^2}\sum_{i,\pm}s_{i,\pm}\bm{\delta}_i$. The new center is $\bm{c'}_0=\left(\bm{c}_0+\eta\Delta\bm{c}_0\right)$, where $\eta$ is the learning rate parameter. The scale parameter is also updated using the gradient $\Delta\sigma=\frac{1}{n\sigma}\sum_{i,\pm}s_{i,\pm}\left(\frac{\bm{\delta}_i^2}{\sigma^2}-1\right)$ and a separate learning rate. Note that we use the same $\sigma$ for all components in the image code and thus do not model a multidimensional Gaussian nor any covariances, different from the general case discussed in \citet{wiesch14}. However, we do update the scale of the search distribution, different from \citet{ilylin18}. We have tried updating separate, independent $\sigma$'s for each component in the image code, but the performance is much worse, presumably because there is too little information to reliably estimate gradients for the second moment.
\setlength{\parindent}{0pt}

\paragraph{Converting image to image codes.}
In several cases, we need to convert an image into an image code in the generative feature space that the generative model takes as input. We used two heuristic methods, ``opt'' and ``ivt''. The ``opt'' method is: Starting with an all-zero image code, the image code is iteratively optimized using backpropagation and gradient descent to minimize the pixel-wise difference between the generated image and the target image. The ``ivt'' method is: The target image is forwarded through CaffeNet, and the fc6 layer activation (post-ReLU activation) is used as the image code, because the generator was originally trained to invert this encoding \citep{dosbro16}.

\paragraph{Hyperparameter optimization}
To obtain a set of reasonably good default hyperparameters, we used a greedy algorithm that maximized performance over a small set of target units by varying one hyperparameter at a time. To keep the computation tractable, we used 12 units total, 3 randomly chosen from the output layer of each of the 4 networks shown in \textbf{Figure \ref{fig:nets}b}. Starting from an educated guess of hyperparameter values, one hyperparameter is chosen at a time. Four \emph{test} values are chosen around the current value with a pre-defined step size, and optimization performance is measured with the test values. The value that yielded the best performance is set as the current value. Then, another hyperparameter is chosen to be varied. The same hyperparameter is not chosen again until all others have been considered once; we call each repeat of all hyperparameters one \emph{round}. If no hyperparameter has been updated in one round, the step size is decreased for the hyperparameter that has not been updated for the longest time. This is repeated until all pre-defined, progressively decreasing step sizes for each parameter have been exhausted. The final best parameter settings are used as the default values.

\paragraph{Stochastic neuron models.} Let $y$ be the activation value of a ConvNet unit to an input image. The activation value corrupted by stochastic noise is $Y\sim\text{Poisson}(\max(0, a))$. To simulate repeated image presentations, a common practice in neuron recording, we simply draw multiple $Y$ values from the same Poisson distribution.

\bibliographystyle{ref}
\bibliography{ref}

\begin{thebibliography}{40}
\providecommand{\natexlab}[1]{#1}
\providecommand{\url}[1]{\texttt{#1}}
\expandafter\ifx\csname urlstyle\endcsname\relax
  \providecommand{\doi}[1]{doi: #1}\else
  \providecommand{\doi}{doi: \begingroup \urlstyle{rm}\Url}\fi

\bibitem[Abbasi-Asl et~al.(2018)Abbasi-Asl, Chen, Bloniarz, Oliver, Willmore,
  Gallant, and Yu]{abbyu18}
Reza Abbasi-Asl, Yuansi Chen, Adam Bloniarz, Michael Oliver, Ben~D.B. Willmore,
  Jack~L. Gallant, and Bin Yu.
\newblock The {DeepTune} framework for modeling and characterizing neurons in
  visual cortex area {V4}.
\newblock \emph{bioRxiv}, 2018.
\newblock \doi{10.1101/465534}.
\newblock URL \url{https://www.biorxiv.org/content/early/2018/11/09/465534}.

\bibitem[Bashivan et~al.(2019 \textit{in press})Bashivan, Kar, and
  DiCarlo]{basdic19}
Pouya Bashivan, Kohitij Kar, and James~J DiCarlo.
\newblock Neural population control via deep image synthesis.
\newblock \emph{Science}, 2019 \textit{in press}.

\bibitem[Brock et~al.(2018)Brock, Donahue, and Simonyan]{brosim18}
Andrew Brock, Jeff Donahue, and Karen Simonyan.
\newblock Large scale {GAN} training for high fidelity natural image synthesis,
  2018.

\bibitem[Bruce et~al.(1981)Bruce, Desimone, and Gross]{brugro81}
Charles Bruce, Robert Desimone, and Charles~G Gross.
\newblock Visual properties of neurons in a polysensory area in superior
  temporal sulcus of the macaque.
\newblock \emph{Journal of neurophysiology}, 46\penalty0 (2):\penalty0
  369--384, 1981.

\bibitem[Carlson et~al.(2011)Carlson, Rasquinha, Zhang, and Connor]{carcha11}
Eric~T Carlson, Russell~J Rasquinha, Kechen Zhang, and Charles~E Connor.
\newblock A sparse object coding scheme in area {V4}.
\newblock \emph{Current Biology}, 21\penalty0 (4):\penalty0 288--293, 2011.

\bibitem[Carter et~al.(2019)Carter, Armstrong, Schubert, Johnson, and
  Olah]{carola19}
Shan Carter, Zan Armstrong, Ludwig Schubert, Ian Johnson, and Chris Olah.
\newblock Activation atlas.
\newblock \emph{Distill}, 2019.
\newblock \doi{10.23915/distill.00015}.
\newblock URL \url{https://distill.pub/2019/activation-atlas}.

\bibitem[Chen et~al.(2017)Chen, Zhang, Sharma, Yi, and Hsieh]{chehsi17}
Pin-Yu Chen, Huan Zhang, Yash Sharma, Jinfeng Yi, and Cho-Jui Hsieh.
\newblock {ZOO}: Zeroth order optimization based black-box attacks to deep
  neural networks without training substitute models.
\newblock In \emph{Proceedings of the 10th ACM Workshop on Artificial
  Intelligence and Security}, pp.\  15--26. ACM, 2017.

\bibitem[Donahue(2014)]{don14}
Jeff Donahue.
\newblock {BVLC Reference CaffeNet}.
\newblock
  \url{https://github.com/BVLC/caffe/tree/master/models/bvlc_reference_caffenet},
  2014.
\newblock Accessed: 2019-04-23.

\bibitem[Dosovitskiy \& Brox(2016)Dosovitskiy and Brox]{dosbro16}
Alexey Dosovitskiy and Thomas Brox.
\newblock Generating images with perceptual similarity metrics based on deep
  networks.
\newblock In D.~D. Lee, M.~Sugiyama, U.~V. Luxburg, I.~Guyon, and R.~Garnett
  (eds.), \emph{Advances in Neural Information Processing Systems 29}, pp.\
  658--666. Curran Associates, Inc., 2016.
\newblock URL
  \url{http://papers.nips.cc/paper/6158-generating-images-with-perceptual-similarity-metrics-based-on-deep-networks.pdf}.

\bibitem[He et~al.(2016)He, Zhang, Ren, and Sun]{hesun16}
Kaiming He, Xiangyu Zhang, Shaoqing Ren, and Jian Sun.
\newblock Identity mappings in deep residual networks.
\newblock \emph{Lecture Notes in Computer Science}, pp.\  630--645, 2016.
\newblock ISSN 1611-3349.
\newblock \doi{10.1007/978-3-319-46493-0_38}.
\newblock URL \url{http://dx.doi.org/10.1007/978-3-319-46493-0_38}.

\bibitem[Hubel \& Wiesel(1962)Hubel and Wiesel]{hubwie62}
David~H Hubel and Torsten~N Wiesel.
\newblock Receptive fields, binocular interaction and functional architecture
  in the cat's visual cortex.
\newblock \emph{The Journal of physiology}, 160\penalty0 (1):\penalty0
  106--154, 1962.

\bibitem[Hung et~al.(2005)Hung, Kreiman, Poggio, and DiCarlo]{hundic05}
Chou~P. Hung, Gabriel Kreiman, Tomaso Poggio, and James~J. DiCarlo.
\newblock Fast readout of object identity from macaque inferior temporal
  cortex.
\newblock \emph{Science}, 310\penalty0 (5749):\penalty0 863--866, 2005.
\newblock ISSN 0036-8075.
\newblock \doi{10.1126/science.1117593}.
\newblock URL \url{https://science.sciencemag.org/content/310/5749/863}.

\bibitem[Ilyas et~al.(2018)Ilyas, Engstrom, Athalye, and Lin]{ilylin18}
Andrew Ilyas, Logan Engstrom, Anish Athalye, and Jessy Lin.
\newblock Black-box adversarial attacks with limited queries and information.
\newblock In Jennifer Dy and Andreas Krause (eds.), \emph{Proceedings of the
  35th International Conference on Machine Learning}, volume~80 of
  \emph{Proceedings of Machine Learning Research}, pp.\  2137--2146,
  Stockholmsmässan, Stockholm Sweden, 10--15 Jul 2018. PMLR.
\newblock URL \url{http://proceedings.mlr.press/v80/ilyas18a.html}.

\bibitem[Jia et~al.(2014)Jia, Shelhamer, Donahue, Karayev, Long, Girshick,
  Guadarrama, and Darrell]{jiadar14}
Yangqing Jia, Evan Shelhamer, Jeff Donahue, Sergey Karayev, Jonathan Long, Ross
  Girshick, Sergio Guadarrama, and Trevor Darrell.
\newblock Caffe: Convolutional architecture for fast feature embedding.
\newblock \emph{arXiv preprint arXiv:1408.5093}, 2014.

\bibitem[Krizhevsky et~al.(2012)Krizhevsky, Sutskever, and Hinton]{krihin12}
Alex Krizhevsky, Ilya Sutskever, and Geoffrey~E Hinton.
\newblock {ImageNet} classification with deep convolutional neural networks.
\newblock In F.~Pereira, C.~J.~C. Burges, L.~Bottou, and K.~Q. Weinberger
  (eds.), \emph{Advances in Neural Information Processing Systems 25}, pp.\
  1097--1105. Curran Associates, Inc., 2012.
\newblock URL
  \url{http://papers.nips.cc/paper/4824-imagenet-classification-with-deep-convolutional-neural-networks.pdf}.

\bibitem[Lee(2017)]{alb14}
Albert Lee.
\newblock Pre-trained caffe models.
\newblock \url{https://github.com/GeekLiB/caffe-model}, 2017.
\newblock Accessed: 2019-04-29.

\bibitem[Malakhova(2018)]{mal18}
K.~Malakhova.
\newblock Visualization of information encoded by neurons in the higher-level
  areas of the visual system.
\newblock \emph{J. Opt. Technol.}, 85\penalty0 (8):\penalty0 494--498, Aug
  2018.
\newblock \doi{10.1364/JOT.85.000494}.
\newblock URL \url{http://jot.osa.org/abstract.cfm?URI=jot-85-8-494}.

\bibitem[Morcos et~al.(2018)Morcos, Raghu, and Bengio]{morben18}
Ari Morcos, Maithra Raghu, and Samy Bengio.
\newblock Insights on representational similarity in neural networks with
  canonical correlation.
\newblock In S.~Bengio, H.~Wallach, H.~Larochelle, K.~Grauman, N.~Cesa-Bianchi,
  and R.~Garnett (eds.), \emph{Advances in Neural Information Processing
  Systems 31}, pp.\  5732--5741. Curran Associates, Inc., 2018.
\newblock URL
  \url{http://papers.nips.cc/paper/7815-insights-on-representational-similarity-in-neural-networks-with-canonical-correlation.pdf}.

\bibitem[Nguyen et~al.(2016)Nguyen, Dosovitskiy, Yosinski, Brox, and
  Clune]{nguclu16}
Anh Nguyen, Alexey Dosovitskiy, Jason Yosinski, Thomas Brox, and Jeff Clune.
\newblock Synthesizing the preferred inputs for neurons in neural networks via
  deep generator networks.
\newblock In \emph{Advances in Neural Information Processing Systems}, pp.\
  3387--3395, 2016.

\bibitem[Olah et~al.(2017)Olah, Mordvintsev, and Schubert]{olasch17}
Chris Olah, Alexander Mordvintsev, and Ludwig Schubert.
\newblock Feature visualization.
\newblock \emph{Distill}, 2017.
\newblock \doi{10.23915/distill.00007}.
\newblock URL \url{https://distill.pub/2017/feature-visualization}.

\bibitem[Olah et~al.(2018)Olah, Satyanarayan, Johnson, Carter, Schubert, Ye,
  and Mordvintsev]{olamor18}
Chris Olah, Arvind Satyanarayan, Ian Johnson, Shan Carter, Ludwig Schubert,
  Katherine Ye, and Alexander Mordvintsev.
\newblock The building blocks of interpretability.
\newblock \emph{Distill}, 2018.
\newblock \doi{10.23915/distill.00010}.
\newblock URL \url{https://distill.pub/2018/building-blocks}.

\bibitem[Pasupathy \& Connor(2002)Pasupathy and Connor]{pascha02}
Anitha Pasupathy and Charles~E Connor.
\newblock Population coding of shape in area {V4}.
\newblock \emph{Nature neuroscience}, 5\penalty0 (12):\penalty0 1332, 2002.

\bibitem[Ponce et~al.(2019 \textit{in press})Ponce, Xiao, Schade, Hartmann,
  Kreiman, and Livingstone]{pxstkm19}
Carlos~R. Ponce, Will Xiao, Peter Schade, Till~S Hartmann, Gabriel Kreiman, and
  Margaret~S Livingstone.
\newblock Evolving images for visual neurons using a deep generative network
  reveals coding principles and neuronal preferences.
\newblock \emph{Cell}, 2019 \textit{in press}.

\bibitem[Portilla \& Simoncelli(2000)Portilla and Simoncelli]{porsim00}
Javier Portilla and Eero~P. Simoncelli.
\newblock A parametric texture model based on joint statistics of complex
  wavelet coefficients.
\newblock \emph{International Journal of Computer Vision}, 40\penalty0
  (1):\penalty0 49--70, Oct 2000.
\newblock \doi{10.1023/A:1026553619983}.
\newblock URL \url{https://doi.org/10.1023/A:1026553619983}.

\bibitem[Rajalingham et~al.(2018)Rajalingham, Issa, Bashivan, Kar, Schmidt, and
  DiCarlo]{rajdic18}
Rishi Rajalingham, Elias~B. Issa, Pouya Bashivan, Kohitij Kar, Kailyn Schmidt,
  and James~J. DiCarlo.
\newblock Large-scale, high-resolution comparison of the core visual object
  recognition behavior of humans, monkeys, and state-of-the-art deep artificial
  neural networks.
\newblock \emph{Journal of Neuroscience}, 38\penalty0 (33):\penalty0
  7255--7269, 2018.
\newblock ISSN 0270-6474.
\newblock \doi{10.1523/JNEUROSCI.0388-18.2018}.
\newblock URL \url{http://www.jneurosci.org/content/38/33/7255}.

\bibitem[Russakovsky et~al.(2015)Russakovsky, Deng, Su, Krause, Satheesh, Ma,
  Huang, Karpathy, Khosla, Bernstein, Berg, and Fei-Fei]{rusli15}
Olga Russakovsky, Jia Deng, Hao Su, Jonathan Krause, Sanjeev Satheesh, Sean Ma,
  Zhiheng Huang, Andrej Karpathy, Aditya Khosla, Michael Bernstein,
  Alexander~C. Berg, and Li~Fei-Fei.
\newblock {ImageNet} large scale visual recognition challenge.
\newblock \emph{International Journal of Computer Vision}, 115\penalty0
  (3):\penalty0 211--252, 2015.
\newblock ISSN 1573-1405.
\newblock \doi{10.1007/s11263-015-0816-y}.
\newblock URL \url{https://doi.org/10.1007/s11263-015-0816-y}.

\bibitem[Schrimpf et~al.(2018)Schrimpf, Kubilius, Hong, Majaj, Rajalingham,
  Issa, Kar, Bashivan, Prescott-Roy, Schmidt, Yamins, and DiCarlo]{schdic18}
Martin Schrimpf, Jonas Kubilius, Ha~Hong, Najib~J. Majaj, Rishi Rajalingham,
  Elias~B. Issa, Kohitij Kar, Pouya Bashivan, Jonathan Prescott-Roy, Kailyn
  Schmidt, Daniel L.~K. Yamins, and James~J. DiCarlo.
\newblock Brain-score: Which artificial neural network for object recognition
  is most brain-like?
\newblock \emph{bioRxiv}, 2018.
\newblock \doi{10.1101/407007}.
\newblock URL \url{https://www.biorxiv.org/content/early/2018/09/05/407007}.

\bibitem[Serre(2019 \textit{in press})]{ser19}
Thomas Serre.
\newblock Deep learning: the good, the bad and the ugly.
\newblock \emph{Annual Review of Vision}, 2019 \textit{in press}.

\bibitem[Simonyan et~al.(2013)Simonyan, Vedaldi, and Zisserman]{simzis13}
Karen Simonyan, Andrea Vedaldi, and Andrew Zisserman.
\newblock Deep inside convolutional networks: Visualising image classification
  models and saliency maps, 2013.

\bibitem[Stanley(2007)]{sta07}
Kenneth~O. Stanley.
\newblock Compositional pattern producing networks: A novel abstraction of
  development.
\newblock \emph{Genetic Programming and Evolvable Machines}, 8\penalty0
  (2):\penalty0 131--162, June 2007.
\newblock ISSN 1389-2576.
\newblock \doi{10.1007/s10710-007-9028-8}.
\newblock URL \url{http://dx.doi.org/10.1007/s10710-007-9028-8}.

\bibitem[Szegedy et~al.(2016)Szegedy, Vanhoucke, Ioffe, Shlens, and
  Wojna]{szewoj16}
Christian Szegedy, Vincent Vanhoucke, Sergey Ioffe, Jon Shlens, and Zbigniew
  Wojna.
\newblock Rethinking the {Inception} architecture for computer vision.
\newblock In \emph{Proceedings of the IEEE conference on computer vision and
  pattern recognition}, pp.\  2818--2826, 2016.

\bibitem[Szegedy et~al.(2017)Szegedy, Ioffe, Vanhoucke, and Alemi]{szeale17}
Christian Szegedy, Sergey Ioffe, Vincent Vanhoucke, and Alexander~A Alemi.
\newblock {Inception-v4}, {Inception-ResNet} and the impact of residual
  connections on learning.
\newblock In \emph{Thirty-First AAAI Conference on Artificial Intelligence},
  2017.

\bibitem[Tang et~al.(2018)Tang, Schrimpf, Lotter, Moerman, Paredes,
  Ortega~Caro, Hardesty, Cox, and Kreiman]{tankre18}
Hanlin Tang, Martin Schrimpf, William Lotter, Charlotte Moerman, Ana Paredes,
  Josue Ortega~Caro, Walter Hardesty, David Cox, and Gabriel Kreiman.
\newblock Recurrent computations for visual pattern completion.
\newblock \emph{Proceedings of the National Academy of Sciences}, 115\penalty0
  (35):\penalty0 8835--8840, 2018.
\newblock ISSN 0027-8424.
\newblock \doi{10.1073/pnas.1719397115}.
\newblock URL \url{https://www.pnas.org/content/115/35/8835}.

\bibitem[Tsao et~al.(2006)Tsao, Freiwald, Tootell, and Livingstone]{tsaliv06}
Doris~Y. Tsao, Winrich~A. Freiwald, Roger B.~H. Tootell, and Margaret~S.
  Livingstone.
\newblock A cortical region consisting entirely of face-selective cells.
\newblock \emph{Science}, 311\penalty0 (5761):\penalty0 670--674, 2006.
\newblock ISSN 0036-8075.
\newblock \doi{10.1126/science.1119983}.
\newblock URL \url{https://science.sciencemag.org/content/311/5761/670}.

\bibitem[Vaziri et~al.(2014)Vaziri, Carlson, Wang, and Connor]{vazcha14}
Siavash Vaziri, Eric~T Carlson, Zhihong Wang, and Charles~E Connor.
\newblock A channel for {3D} environmental shape in anterior inferotemporal
  cortex.
\newblock \emph{Neuron}, 84\penalty0 (1):\penalty0 55--62, 2014.

\bibitem[Walker et~al.(2018)Walker, Sinz, Froudarakis, Fahey, Muhammad, Ecker,
  Cobos, Reimer, Pitkow, and Tolias]{waltol18}
Edgar~Y. Walker, Fabian~H. Sinz, Emmanouil Froudarakis, Paul~G. Fahey, Taliah
  Muhammad, Alexander~S. Ecker, Erick Cobos, Jacob Reimer, Xaq Pitkow, and
  Andreas~S. Tolias.
\newblock Inception in visual cortex: in vivo-silico loops reveal most exciting
  images.
\newblock \emph{bioRxiv}, 2018.
\newblock \doi{10.1101/506956}.
\newblock URL \url{https://www.biorxiv.org/content/early/2018/12/28/506956}.

\bibitem[Wierstra et~al.(2014)Wierstra, Schaul, Glasmachers, Sun, Peters, and
  Schmidhuber]{wiesch14}
Daan Wierstra, Tom Schaul, Tobias Glasmachers, Yi~Sun, Jan Peters, and
  J\"{u}rgen Schmidhuber.
\newblock Natural evolution strategies.
\newblock \emph{Journal of Machine Learning Research}, 15:\penalty0 949--980,
  2014.
\newblock URL \url{http://jmlr.org/papers/v15/wierstra14a.html}.

\bibitem[Yamane et~al.(2008)Yamane, Carlson, Bowman, Wang, and
  Connor]{yamcha08}
Yukako Yamane, Eric~T Carlson, Katherine~C Bowman, Zhihong Wang, and Charles~E
  Connor.
\newblock A neural code for three-dimensional object shape in macaque
  inferotemporal cortex.
\newblock \emph{Nature neuroscience}, 11\penalty0 (11):\penalty0 1352, 2008.

\bibitem[Yamins et~al.(2014)Yamins, Hong, Cadieu, Solomon, Seibert, and
  DiCarlo]{yamdic14}
Daniel L.~K. Yamins, Ha~Hong, Charles~F. Cadieu, Ethan~A. Solomon, Darren
  Seibert, and James~J. DiCarlo.
\newblock Performance-optimized hierarchical models predict neural responses in
  higher visual cortex.
\newblock \emph{Proceedings of the National Academy of Sciences}, 111\penalty0
  (23):\penalty0 8619--8624, 2014.
\newblock ISSN 0027-8424.
\newblock \doi{10.1073/pnas.1403112111}.
\newblock URL \url{https://www.pnas.org/content/111/23/8619}.

\bibitem[Zhou et~al.(2014)Zhou, Lapedriza, Xiao, Torralba, and Oliva]{zhooli14}
Bolei Zhou, Agata Lapedriza, Jianxiong Xiao, Antonio Torralba, and Aude Oliva.
\newblock Learning deep features for scene recognition using {Places} database.
\newblock In \emph{Proceedings of the 27th International Conference on Neural
  Information Processing Systems - Volume 1}, pp.\  487--495, Cambridge, MA,
  USA, 2014. MIT Press.
\newblock URL \url{http://dl.acm.org/citation.cfm?id=2968826.2968881}.

\end{thebibliography}

\section{Supplementary Material}
\setcounter{figure}{0}
\renewcommand{\thefigure}{S\arabic{figure}}
\setcounter{table}{0}
\renewcommand{\thetable}{S\arabic{table}}

\subsection{Random exploration of stimulus space}
\label{natmax}
\begin{figure}
  \centering
  \includegraphics[width=16.5 cm]{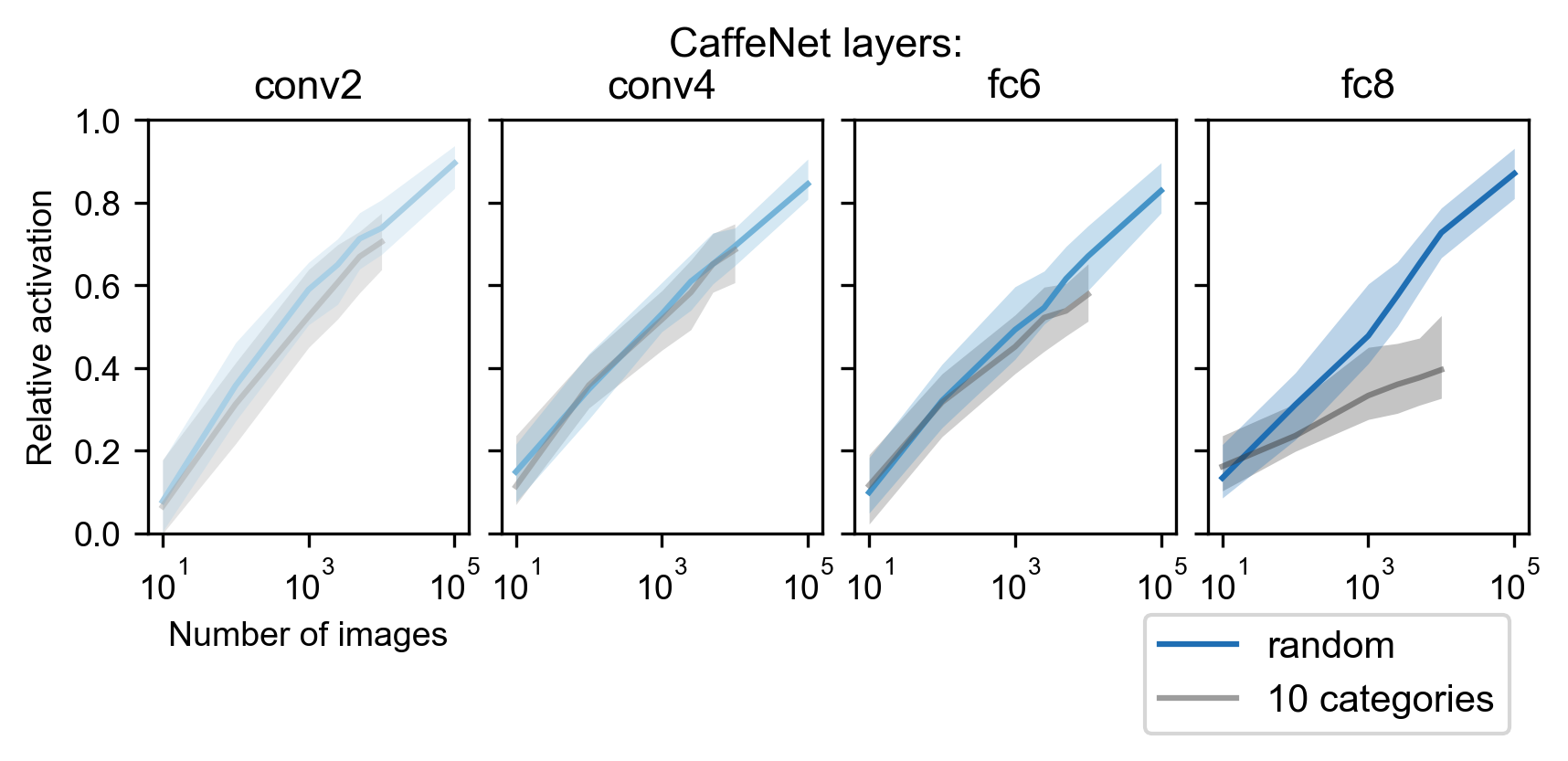}
  \caption{
    \textbf{Expected maximum relative activation in response to natural images with random sampling.} We measured the max relative activation expected in two random sampling schemes. "Random" refers to picking a given number of images randomly from the imageNet dataset (blue). "Category" refers to first randomly picking 10 categories out of the 1000 imageNet categories and then picking a given number of images randomly from those categories so that the total number of images is the one indicated on the x-axis (gray). We considered 4 layers from the CaffeNet architecture. Lines indicates the median relative activation (largest activation divided by the largest activation for all imageNet images). Shading indicates the 25th- to 75th-percentiles among 100 random units per layer.
  }
  \label{fig:natmax}
\end{figure}

A common approach in neuroscience for exploring neuronal selectivity has been to use arbitrarily selected images, often from a limited number of categories. As a point of comparison, we randomly sampled $n$ images either 1) from all of ImageNet, or 2) from 10 categories randomly selected from the 1,000 training categories in ImageNet ($n/10$ exemplars per category). We measured the activation value to each image for units in different layers of CaffeNet, and calculated the maximum observed relative activation (activation value normalized by that of the best image in all $>1.4$ M images in ImageNet). As expected, the maximum observed relative activation increased with $n$ without any optimization protocol, but only did so slowly with a near-logarithmic speed of growth (\textbf{Figure \ref{fig:natmax}}). Moreover, for higher layers (e.g., fc8), selecting the images from only 10 categories yielded significantly worse results than selecting the images randomly, which we hypothesize is because the small number of categories sets a bottleneck on the diversity of high-level features represented.

\subsection{Expressiveness and searchability}
\label{searchability}
\begin{figure}
  \centering
  \includegraphics[width=16.5 cm]{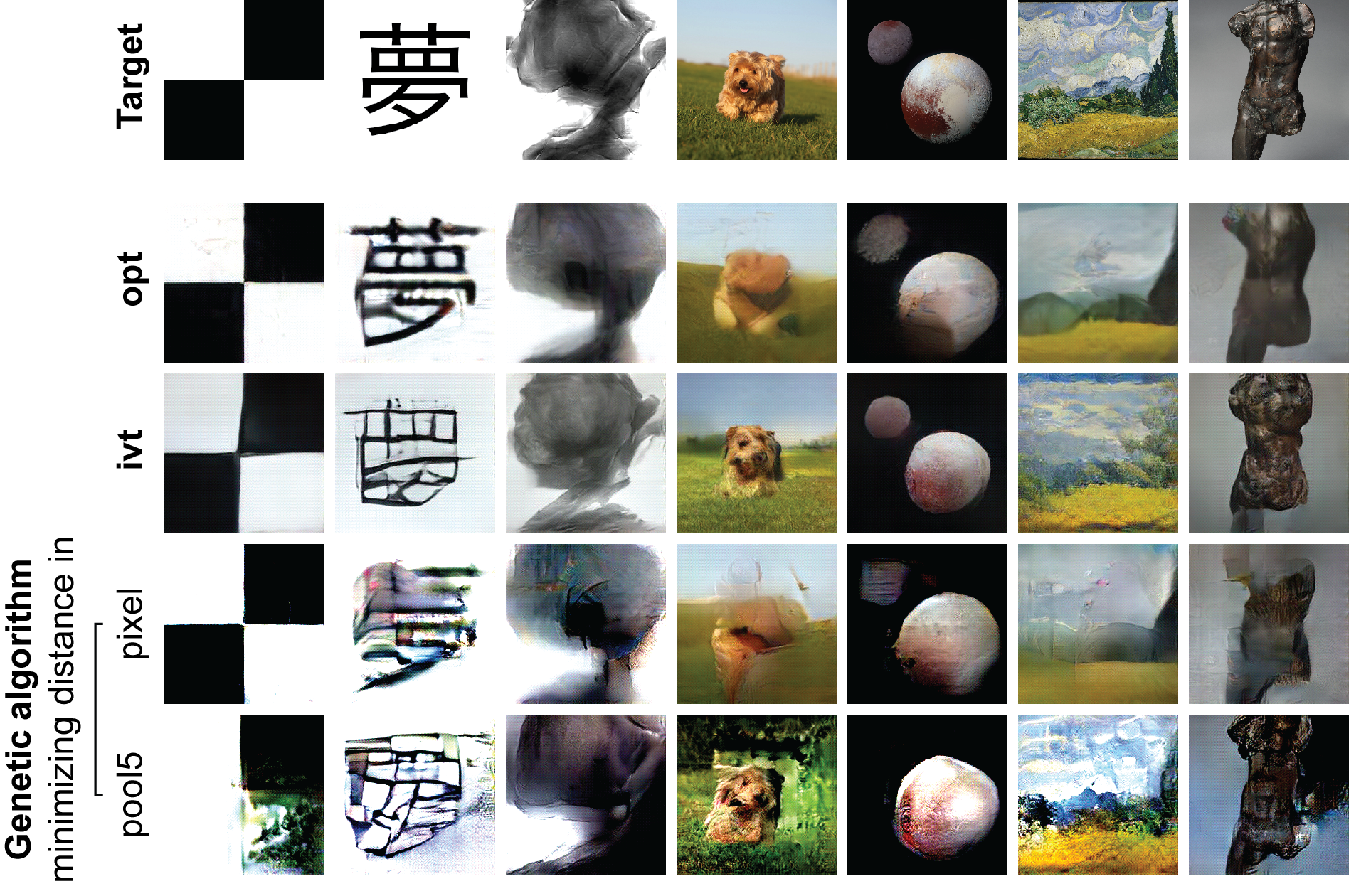}
  \caption{
    \textbf{The generative network can approximately represent arbitrary images, and XDream can uncover these images using only scalar distance as a loss function.} The generative network is challenged to synthesize arbitrary target images (row 1) using one of two encoding methods, ``opt'' (row 2) and ``ivt'' (row 3; \textbf{Methods}). In addition, we tested whether XDream can discover the ground truth target image efficiently using the genetic algorithm. For each target image, we constructed a loss function using a scalar distance between the target image and any test image, and used XDream to minimize the loss function. The distance is the average squared difference between representations either in pixel space (row 4) or CaffeNet pool5 space (row 5). The distance metric is convex, so there is a well-defined global optimum by construction. XDream is only given 10,000 test image presentations. The source of the target images are as follows: the leftmost 2 images are rendered; the third image is texture synthesized as described in \citet{porsim00}; the fourth image is from the ImageNet test set; the rightmost 3 images are public domain images from NASA and The Metropolitan Museum of Art.
  }
  \label{fig:groundtruth}
\end{figure}

Is XDream limited in what images it can find? We have discussed this issue in \citet{pxstkm19}, but the question is relevant here so we reproduce the analysis with slightly different data. Because we optimize in the image code space of a generative network, a first constraint is the co-domain of the generative network. It is hard to quantify what fraction of possible images is represented by a generative network. We attempted to qualitative assess the expressiveness of the generative network by challenging it to synthesize diverse, arbitrarily selected target images (\textbf{Figure \ref{fig:groundtruth}}). The results indicate that the generative network is able to encode, at least approximately, all the tested target images.

However, the generative network not only has to represent diverse images, it also has to be efficiently (in a reasonable number of steps) searchable by the optimization algorithm. This question depends on both the optimization algorithm and the loss function guiding the search. We used a simple loss function that is just the mean square difference between the target image and any input image, computed either with the pixel representation or with the CaffeNet pool5 layer representation of the image. In both cases, XDream is able to uncover an image qualitatively resembling the original (\textbf{Figure \ref{fig:groundtruth}}). At least part of the remaining difference could be attributed to the loss function: Pixel-wise loss is known to lead to excessive smoothing, and pool5 loss is expected to lose some features and spatial information due to pooling operations and ReLU activations in preceding layers.

These results suggest that XDream is able to efficiently search a large, expressive stimulus space.

\end{document}